\documentclass[preprint]{vgtc}               

\makeatletter
\renewcommand{\vgtc@preprinttext}{%
  \parbox{0.9\textwidth}{%
    $\copyright$ \the\year~IEEE. This is the author's version of the 1st Place winning report from the VISxGenAI Challenge at IEEE VIS 2025. The final version is available on the workshop website at: \href{https://visxgenai.github.io/subs-2025/3765/3765-doc.pdf}{\color{blue}visxgenai.github.io}.
  }%
}
\makeatother

\graphicspath{{figures/}{pictures/}{images/}{./}} 

\usepackage{times}                     

\usepackage{tikz}
\usepackage[dvipsnames]{xcolor} 
\definecolor{nodeblue}{RGB}{20, 82, 251} 

\newcommand*\nodecircle[1]{\tikz[baseline=-0.7ex]{
    \node[shape=circle, fill=nodeblue, text=white, inner sep=1.5pt, font=\sffamily\bfseries\tiny] (char) {#1};}
}

\usepackage{mathptmx}                  

\onlineid{3765}

\vgtccategory{Research}

\vgtcinsertpkg

\title{A Composable Agentic System for Automated Visual Data Reporting}

\author{
Péter Ferenc Gyarmati\thanks{e-mail: \href{mailto:peter.gyarmati@univie.ac.at}{peter.ferenc.gyarmati@univie.ac.at}}\\
    \scriptsize University of Vienna
\and
Dominik Moritz\thanks{e-mail: \href{mailto:domoritz@cmu.edu}{domoritz@cmu.edu}}\\
    \parbox{1.4in}{\scriptsize \centering Carnegie Mellon University}
\and
Torsten Möller\thanks{e-mail: \href{mailto:torsten.moeller@univie.ac.at}{torsten.moeller@univie.ac.at}}\\
    \scriptsize University of Vienna
\and
Laura Koesten\thanks{e-mail: \href{mailto:laura.koesten@mbzuai.ac.ae}{laura.koesten@mbzuai.ac.ae}}\\
    \parbox{2in}{\scriptsize \centering MBZUAI \\ Austrian Institute of Technology \\ University of Vienna}
}

\teaser{
  \centering
  \includegraphics[width=1.0\linewidth]{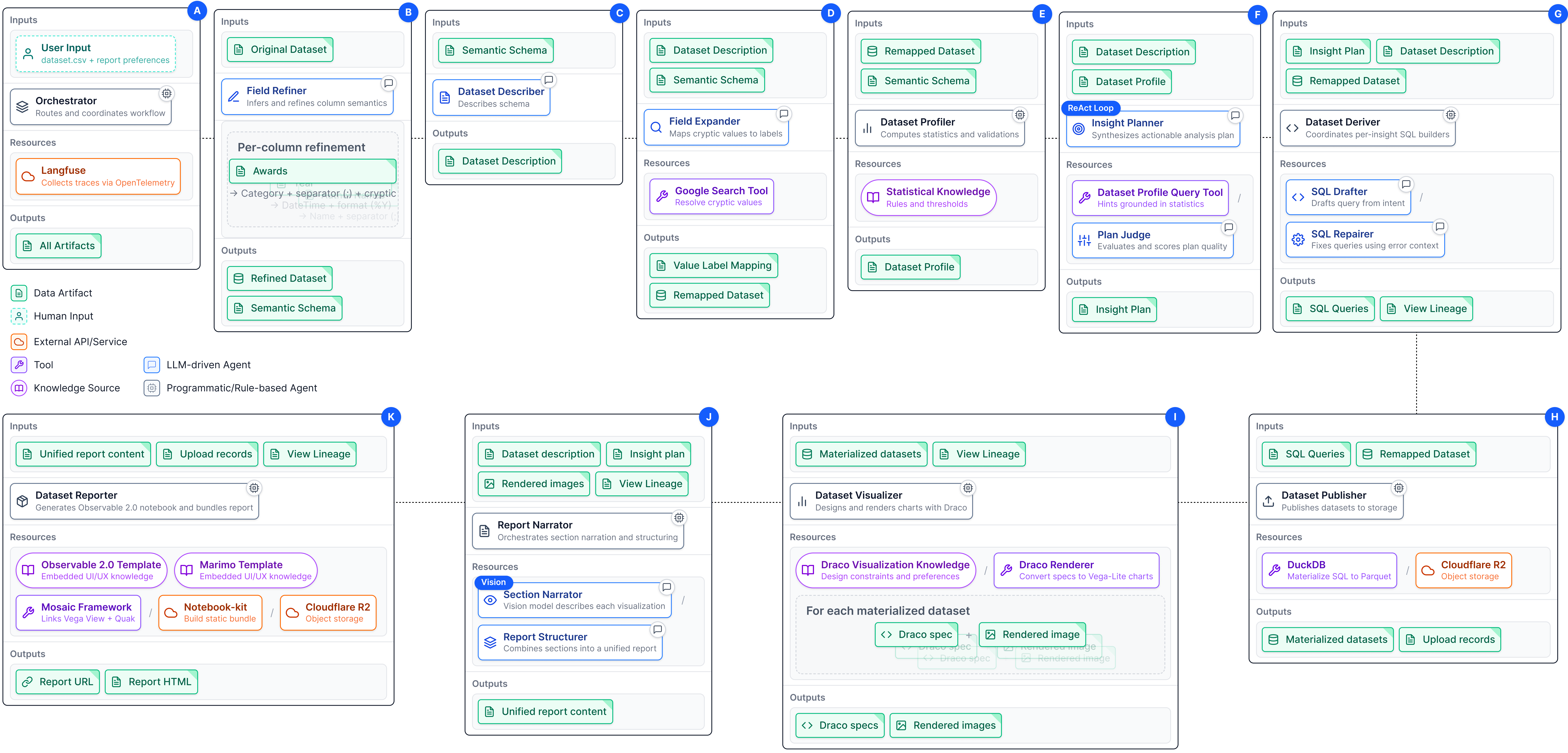}
  \caption{The architecture of our agentic system, a multi-stage pipeline that deconstructs report generation. This hybrid architecture externalizes critical logic to deterministic components like the \texttt{Dataset Profiler} and the rule-based \texttt{Dataset Visualizer} powered by Draco~\cite{Yang:2023:Draco2}. The system's granular artifacts are assembled into a dual output: an interactive Observable~\cite{Observable:2025:NBKit} report with Mosaic~\cite{Heer:2024:Mosaic} for engaging reader exploration, and executable Marimo~\cite{Marimo:2023:Notebooks} notebooks that provide deep, per-insight provenance.}
\label{fig:teaser}
}

\abstract{
To address the brittleness of monolithic AI agents, our prototype for automated visual data reporting explores a Human-AI Partnership model. Its hybrid, multi-agent architecture strategically externalizes logic from LLMs to deterministic modules, leveraging the rule-based system Draco~\cite{Yang:2023:Draco2} for principled visualization design. The system delivers a dual-output: an interactive Observable~\cite{Observable:2025:NBKit} report with Mosaic~\cite{Heer:2024:Mosaic} for reader exploration, and executable Marimo~\cite{Marimo:2023:Notebooks} notebooks for deep, analyst-facing traceability. This granular architecture yields a fully automatic yet auditable and steerable system, charting a path toward a more synergistic partnership between human experts and AI. For reproducibility, our implementation and examples are available at \href{https://peter-gy.github.io/VISxGenAI-2025/}{peter-gy.github.io/VISxGenAI-2025}.
} 


\begin{document}


\firstsection{Introduction}

\maketitle

The proliferation of autonomous agents is reshaping data analysis and some proclaim---despite no evidence---that they will replace human analysts. We explore a \textbf{Human-AI Partnership Model}, where the AI serves as a powerful assistant that produces a complete, initial analysis, and the human expert acts as a strategist and supervisor. The goal is to move beyond the paradigm of a brittle, black-box agent toward a more explainable, steerable, and collaborative workflow.

This partnership model is motivated by a core challenge in data analysis. Even a technically flawless agent may fail if it perfectly executes the \textit{wrong task}. Human intent is often ambiguous at the outset and evolves as insights emerge. A minor misalignment can render an agent's entire output useless, forcing costly and unstable regeneration that erodes user trust and agency. Genuine human–AI partnership requires systems explicitly designed to address this challenge: providing auditable, high-quality outputs while enabling domain experts to iteratively refine them.
To investigate this model, we developed a prototype agentic system for automated visual data reporting\footnote{Our workshop challenge submission is available \href{https://peter-gy.github.io/VISxGenAI-2025/reports/vispub-submission/index.html}{online}.}. The design of our implementation was guided by four high-level principles, adapting established guidelines for human-AI interaction~\cite{Amershi:2019:HAIGuidelines} to prioritize human agency alongside automation~\cite{Heer:2019:AgencyPlus}.

\noindent\textbf{Explainability \& Trust.} For an expert to trust and work with an AI, the system's reasoning should be transparent, and its data transformations should be fully traceable to their source. \textbf{Composability \& Modularity.} Instead of relying on a single model's embedded knowledge, a robust system should be highly composable. It should leverage specialized tools for established best practices, externalizing critical knowledge into interpretable, often deterministic modules. This design improves future-readiness, as components can be updated without redesigning the core logic. \textbf{Granularity \& Iterative Outputs.} The analytical process is iterative. Its outputs should therefore be granular, deconstructed artifacts rather than an indivisible monolith. This enables ``surgical modification'' of a single component without a full, resource-intensive regeneration, supporting a more efficient and sustainable workflow. \textbf{Interoperability via Open Standards.} To facilitate seamless human intervention, a system should ``speak the same language'' as the expert community. Using common open standards makes its outputs useful and editable within a user's existing toolchain, facilitating collaboration.

\section{System Design and Implementation}
Our prototype is a multi-agent system implemented using DSPy~\cite{Khattab:2023:DSPy}. As \autoref{fig:teaser} shows, the main \texttt{Orchestrator}~\nodecircle{A} coordinates a workflow of specialized, hybrid agents that combine LLM reasoning with deterministic logic. It accepts as input a dataset URI and high-level user intent, such as the overall goal and desired length of the report. To ensure full audibility, the \texttt{Orchestrator} collects and stores detailed execution traces of all agent interactions in Langfuse~\cite{Langfuse:2025:LLMPlatform}, an external observability service.

\noindent\textbf{1. Data Understanding.} The pipeline begins with data preparation: the \texttt{Field Refiner}~\nodecircle{B} cleans the data and infers column semantics before the \texttt{Dataset Describer}~\nodecircle{C} and \texttt{Field Expander}~\nodecircle{D}, the latter augmented with a web search tool, create a semantic schema and resolve cryptic codes (e.g., `BP` to `Best Paper Award`, `C` to `Conference Paper`). This phase culminates in the programmatic \texttt{Dataset Profiler}~\nodecircle{E}, which computes a comprehensive statistical profile. This externalization of statistical truth into a deterministic module is a core tenet of our \textbf{Composability} principle.

\noindent\textbf{2. Analysis and Materialization.} With a clean dataset, the \texttt{Insight Planner}~\nodecircle{F} operates within a Reasoning and Acting (ReAct) loop~\cite{Yao:2023:ReAct}, using a dedicated tool to retrieve hints from the statistical profile and ground its planning in data. For each planned insight, the \texttt{Dataset Deriver}~\nodecircle{G} then drafts and, if necessary, repairs a DuckDB SQL query~\cite{Raasveldt:2019:DuckDB}, also outputting semantic roles for the generated fields. Adhering to our \textbf{Granularity} principle, the programmatic \texttt{Dataset Publisher}~\nodecircle{H} executes each query and stores the resulting materialized dataset as a versioned Apache Parquet file in Cloudflare R2, an S3-compatible object store~\cite{ApacheParquet, CloudflareR2}.

\noindent\textbf{3. Visualization.} We avoid monolithic LLM-based chart design, a token-intensive approach whose recommendations only moderately align with human preferences and remain widely inconsistent~\cite{Wang:2025:DracoGPT}. Instead, we use LLM-generated metadata from prior stages to inform a deterministic visualization process. The semantic roles from the \texttt{Dataset Deriver}~\nodecircle{G} and the analytical task from the \texttt{Insight Planner}~\nodecircle{F} are used to programmatically construct a partial Draco specification that defines \textit{what} to visualize~\cite{Yang:2023:Draco2}. The \texttt{Dataset Visualizer}~\nodecircle{I} then delegates the task of determining \textit{how} to visualize it to Draco's rule-based solver. Based on its computational analysis of the data's properties (e.g., cardinality, skew), Draco can recommend complex, appropriate charts like a faceted scatterplot with a symmetric logarithmic scale--a nuanced design decision difficult for a pure LLM approach. This provides another advantage: if a recommendation is suboptimal, its logic is transparent. We can systematically improve future results by adjusting Draco's knowledge base, a key benefit of a white-box system.

\noindent\textbf{4. Reporting and Publication.} The final stage packages all artifacts for human interaction. A vision-language model, coordinated by \texttt{Report Narrator}~\nodecircle{J}, generates textual descriptions for each base64-encoded chart image and structures them into a cohesive narrative. The \texttt{Dataset Reporter}~\nodecircle{K} then assembles these with granular artifacts produced by upstream agents. Because all these components are decoupled from presentation, they are modality-agnostic and can be composed into diverse formats, from an interactive web report to a static PDF. Our system demonstrates this flexibility by programmatically constructing a dual output from these same artifacts: JavaScript source for an interactive Observable 2.0 report~\cite{Observable:2025:NB2}, which a self-hosted Node.js service~\cite{NodeJS, Observable:2025:NBKit, Gyarmati:2025:NBBuilder} builds into a static website, and a Python-based executable Marimo notebook~\cite{Marimo:2023:Notebooks} for deep, analyst-facing traceability. The Observable report uses Mosaic~\cite{Heer:2024:Mosaic} to link each visualization with its Quak data table~\cite{Manz:2025:Quak}, an approach inspired by narrative visualization~\cite{Heer:2010:NarrativeVis} that combines the AI's authored insights with interactive reader-driven exploration. The Marimo notebook, in turn, serves as an executable trace for a single insight, revealing the precise data transformation and visualization logic. This dual-output system provides distinct entry points for expert interaction and targeted modification.

By deconstructing the workflow into well-scoped tasks, our modular design allows for a ``right-size'' model strategy~\cite{Belcak:2025:SLM}. Smaller, faster models like Gemini 2.5 Flash~\cite{Gemini25Flash} handle the majority of tasks, reducing overall latency, while the more powerful Gemini 2.5 Pro~\cite{Gemini25Pro} is reserved for the most demanding functions: generating complex SQL~\nodecircle{G} and authoring descriptive visual narratives~\nodecircle{J}. This design makes such systems more future-ready, allowing models to be swapped individually and supporting the use of open-weight, self-hosted alternatives.

\section{Discussion}
Testing on eight diverse datasets demonstrates our system's generalizability, producing meaningful reports without any domain-specific tuning\footnote{All discussed reports are available for exploration \href{https://peter-gy.github.io/VISxGenAI-2025/gallery}{here}. \label{fn:reports}}. The value of our Human-AI Partnership model is best illustrated by the distinct workflows it enables: systemic steerability for the analyst creating the report, and open-ended exploration for the reader consuming it. 
\noindent\textbf{Steerability for the Analyst.} An early report featured a weak insight, faceting a scatterplot by the highly skewed \texttt{GraphicsReplicabilityStamp} field\textsuperscript{\ref{fn:reports}}. Instead of fragile prompt engineering, our architecture supports a durable fix. Using execution traces, an analyst can pinpoint that the \texttt{Insight Planner}~\nodecircle{F} was misled by a hint from its \texttt{Dataset Profile Query Tool}. The analyst then intervenes by modifying the tool's logic, adding a rule to only suggest fields for segmentation that exceed a statistical diversity threshold. This dataset-agnostic refinement permanently improves the system's knowledge base, preventing similar flawed insights in all future runs.
\noindent\textbf{Explorability for the Reader.} The partnership extends to the reader. Our reports use Mosaic~\cite{Heer:2024:Mosaic} to enable exploration beyond the AI's authored narrative. A static chart of downloads vs. citations is enhanced with a linked data table; readers can answer complex ad-hoc questions (e.g., ``What about VIS papers in the last five years?'') by cross-filtering with attributes like conference and year from the linked data table, which are \textit{not} encoded in the main visualization.

This dual capability--deep, systemic refinement for the analyst and open-ended exploration for the reader--embodies our design philosophy. A successful partnership stems not from demanding perfection from the AI, but from ensuring its outputs are both deeply auditable and adaptable for the expert and flexibly explorable for the reader.

\section{Conclusion}
Our prototype demonstrates a Human-AI Partnership model where the agent acts as a transparent and auditable assistant, not a perfect oracle. Our design instantiates this philosophy through a curated, interoperable toolchain. We leverage Draco for principled, white-box visualization design. The final outputs then serve two distinct purposes: an interactive Observable report powered by Mosaic for reader-driven exploration, and deep, executable Marimo notebooks that provide the analyst with full traceability. By providing the full report source code, this granular architecture grants experts agency for targeted refinement. Designing for human agency is how we can transform AI into a genuine partner for trustworthy data-driven discovery.

\bibliographystyle{abbrv-doi}

\bibliography{template}
\end{document}